\documentclass[aps, prb,twocolumn,superscriptaddress]{revtex4-2}

\usepackage{amsmath,amssymb}
\usepackage{graphicx}
\usepackage{dcolumn}
\usepackage{bm}
\usepackage[utf8]{inputenc} \usepackage{textcomp}
\begin{document}

\title{Emergent exchange bias in ultra-thin La$_{0.67}$Sr$_{0.33}$MnO$_3$ films driven by ferro- antiferromagnetic phase coexistence}

\author{I.~García-Manuz}
\affiliation{Departamento de Física de la Materia Condensada. Condensed Matter Physics Center (IFIMAC), Universidad Autónoma de Madrid, Campus de Cantoblanco, 28049 Madrid, Spain}
\author{G.~Caballero}
\affiliation{IMDEA Nanociencia, C/ Faraday 9, 28049 Madrid, Spain.}
\author{J. L. F.~Cuñado}
\affiliation{IMDEA Nanociencia, C/ Faraday 9, 28049 Madrid, Spain.}
\author{M.~Romera}
\affiliation{Departamento de Física de Materiales. Facultad de Física, Universidad Complutense de Madrid, 28040 Madrid , Spain}
\author{M.~Menghini}
\affiliation{IMDEA Nanociencia, C/ Faraday 9, 28049 Madrid, Spain.}
\author{C.~León}
\affiliation{Departamento de Física de Materiales. Facultad de Física, Universidad Complutense de Madrid, 28040 Madrid , Spain}
\author{J.~Santamaría}
\affiliation{Departamento de Física de Materiales. Facultad de Física, Universidad Complutense de Madrid, 28040 Madrid , Spain}
\author{J.~Camarero}
\affiliation{Departamento de Física de la Materia Condensada. Condensed Matter Physics Center (IFIMAC), Universidad Autónoma de Madrid, Campus de Cantoblanco, 28049 Madrid, Spain}
\author{P.~Perna}
\email{ paolo.perna@imdea.org}
\affiliation{IMDEA Nanociencia,  C/ Faraday 9, 28049 Madrid, Spain.}
\affiliation{Unidad de Nanomateriales Avanzados, IMDEA-Nanociencia, Unidad Asociada al CSIC por el ICMM.}
\author{F. Ajejas}
\email{ fernando.ajejas@imdea.org}
\affiliation{IMDEA Nanociencia, C/ Faraday 9, 28049 Madrid, Spain.}
\affiliation{Unidad de Nanomateriales Avanzados, IMDEA-Nanociencia, Unidad Asociada al CSIC por el ICMM.}

\date{\today}

\begin{abstract}
Ultra-thin La$_{0.67}$Sr$_{0.33}$MnO$_3$ (LSMO) films are generally regarded as single-phase ferromagnets (FM), yet their reduced dimensionality enhances the impact of oxygen stoichiometry and local structural distortions. Here we demonstrate that LSMO layers with thicknesses between 3 and 17~nm develop a robust and thickness-independent exchange bias (EB) despite the absence of an engineered FM$|$antiferromagnetic (AFM) interface. Angular magneto-optical Kerr effect measurements reveal a reproducible hysteresis loop shift that reverses sign upon 180$^\circ$ rotation, confirming its intrinsic origin. X-ray photoelectron spectroscopy uncovers a thickness-independent Mn$^{2+}$ fraction ($\sim$20\%), evidencing the presence of oxygen-deficient reduced regions embedded within the Mn$^{3+}$/Mn$^{4+}$ ferromagnetic matrix. A simple interfacial-exchange model indicates that only a sub-percent fraction of AFM  (or frustrated spin‑ice‑like) regions is required to generate the observed EB. These results demonstrate that ultra-thin LSMO can spontaneously host internal FM$|$AFM interfaces driven by oxygen deficiency, revealing an emergent route to exchange bias in nominally single-phase manganite films.
\end{abstract}

\maketitle

\section{Introduction}

Perovskite manganites such as La$_{1-x}$Sr$_x$MnO$_3$ (LSMO) exhibit a rich phase diagram arising from the competition between double exchange, superexchange, Jahn--Teller distortions, and strong electron--lattice coupling \cite{Zener1951, Anderson1955, Millis1995, Coey1999, Tokura2000, Dagotto2001}. Their ground state can be finely tuned through Sr doping \cite{chaluvadi2022}, epitaxial strain \cite{chaluvadi2020}, and oxygen stoichiometry \cite{trappen2019}, enabling control over ferromagnetic (FM), antiferromagnetic (AFM), metallic, insulating, and charge/orbital-ordered phases. This tunability extends to anisotropic transport \cite{Perna2017}, strain-driven metal--insulator transitions \cite{tebano2006}, and optoelectronic \cite{navarro2023} responses. Moreover, oxygen vacancy migration can induce nanoscale phase separation and resistive switching phenomena \cite{salev2021}, making LSMO relevant for memristive and neuromorphic concepts \cite{lopez2023}. Owing to its robust metallic ferromagnetism and high spin polarization at room temperature, the composition La$_{0.67}$Sr$_{0.33}$MnO$_3$ is widely used as a benchmark in oxide spintronics and multifunctional heterostructures.

The exchange bias (EB) effect, classically observed in FM$|$AFM bilayers \cite{nogues1999}, provides a unidirectional anisotropy \cite{jimenez2009, Perna2012, perna2015, diez2022} essential for spin-valves, spin-filters or for electrical manipulation \cite{ajejas2023}. Beyond engineered heterostructures using $3d$ transition metals, EB has also been reported with complex oxides, typically driven by non-stoichiometry, disorder, or strain-induced phase separation \cite{Niebieskikwiat2005, Dong2008, Wu2020, Zhang2021, Zhang2024}. In LSMO-based heterostructures, exchange bias can arise from FM$|$AFM engineered interfaces, and also due to strain, cation disorder, or interfacial chemical reconstruction promoting the formation of nanoscale AFM regions \cite{Wu2020, Ziese2010, Zhang2021, Bibes2008}. However, EB in nominally single-phase ultra-thin LSMO films remains largely unexplored, despite the fact that finite-size effects and oxygen non-stoichiometry become increasingly prominent at the few-nanometer scale. 

Here, we investigate LSMO ultra-thin films  with thicknesses below 20\,nm using  angular temperature-variable magneto-optical Kerr effect (MOKE) \cite{cunado2015}. We found a non-negligible and reproducible EB in all films, confirmed by EB-sign reversal upon sample rotation, ruling out trivial instrumental artifacts. X-ray photoelectron spectroscopy (XPS) analysis shows the presence of Mn$^{2+}$ fraction ($\sim$20\%) independent of thickness, pointing to oxygen-deficiency-induced reduced regions embedded within a ferromagnetic matrix.
We propose that these Mn$^{2+}$-rich regions behave as AFM, frustrated, or even spin‑ice‑like clusters that couple to the surrounding FM network, generating an internal AFM$|$FM interface and giving rise to the observed EB. This work demonstrates that ultrathin LSMO films can spontaneously host interfacial magnetic anisotropies typically associated with engineered FM$|$AFM heterostructures.

\section{Methods}

Epitaxial ultrathin La$_{0.67}$Sr$_{0.33}$MnO$_3$ (LSMO) films with nominal thicknesses of 3, 7, and 17 nm were grown on (001)-oriented SrTiO$_3$ (STO) single crystal substrates by high-temperature sputtering. 
The deposition was carried out at 1200\,K under a pure oxygen atmosphere with a total pressure of 3.2\,mbar, followed by annealing process at 1200\,K in pure oxygen at 1\,atm during 1\,h to ensure full oxidation and stabilize the epitaxial structure. 

Structural characterization was performed using high-resolution x-ray diffraction (XRD), reflectivity (XRR)  and reciprocal space mapping (RSM) around asymmetric reflections of both film and substrate to assess epitaxial quality, strain state, and lattice parameters. High crystalline quality and coherent epitaxy were confirmed by XRD measurements. XPS was carried out using a monochromatic Al K$_{\alpha}$ source (1486.6 eV) and spherical analyzer. The core levels were analyzed following standard background subtraction and peak-fitting procedures. All spectra were charge-corrected by aligning the valence-band edge to the Fermi level.

MOKE measurements were conducted in the longitudinal geometry using a 15 mW diode laser system ($\lambda$= 442 nm). The sample was mounted on a motorized rotation stage into a variable-temperature cryostat with optical access, allowing full 360$^\circ$ in-plane rotation with angular steps of 1$^\circ$. Hysteresis loops were recorded at each angle. To verify the intrinsic nature of the exchange bias, additional angular scans were performed after reversing the sample orientation and after demagnetizing the coils. Magnetization measurements as a function of temperature [$M(T)$] and magnetic field were performed using a SQUID magnetometer, with the magnetic field applied in-plane. Electrical transport, current vs voltage ($I-V$) sweeps as a function of temperature, was measured in a four-probe configuration using an He closed-cycle cryostat.

\section{Experimental}

Three ultrathin La$_{0.67}$Sr$_{0.33}$MnO$_3$ films were structurally characterized with high-resolution XRR and XRD measurements shown in Figure~1 a--c) and Figure~1 d--f), respectively. Open circles represent the experimental diffraction profiles, while the solid lines correspond to the fits obtained from the model. A clear and systematic reduction of the oscillation period is observed as the film thickness increases. This monotonic evolution confirms that the nominal layer thicknesses are accurately reproduced in the structural model and that the films maintain coherent growth across the investigated thickness range. The 3\,nm and 7\,nm films exhibit fully strained growth on SrTiO$_3$ (001), with the LSMO (00\emph{l}) reflection nearly overlapping the substrate peak, consistent with coherent epitaxy in the ultra-thin regime. 
In contrast, the 17\,nm film shows a clearly resolved LSMO peak at $2\theta \approx$~47.4$^\circ$, indicating partial strain relaxation and a more bulk-like out-of-plane lattice parameter.
Reciprocal space maps, provided in the Supplementary Information, reveal well-defined asymmetric reflections from both film and substrate, confirming high epitaxial quality and allowing extraction of the in-plane and out-of-plane lattice parameters. 

\begin{figure} 
\centering
\includegraphics[width=\columnwidth]
{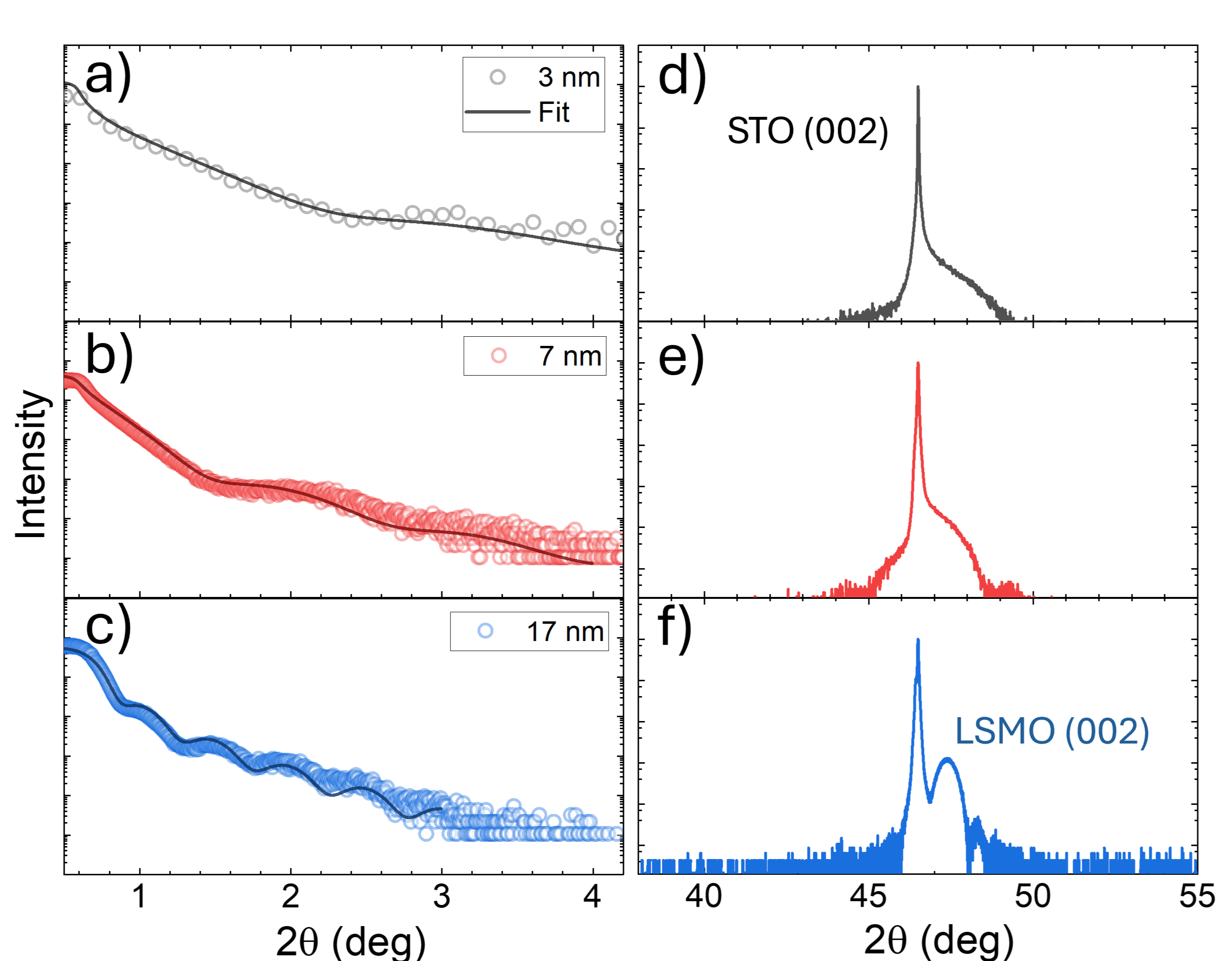}
\caption{ a–c) X-ray reflectivity (XRR) curves for La$_{0.67}$Sr$_{0.33}$MnO$_3$ films with thicknesses of 
3 nm (black), 7 nm (red), and 17 nm (blue). Open circles correspond to experimental data and solid lines to the fitted profiles. 
d–f) High-resolution X-ray diffraction (XRD) scans around the LSMO (00\emph{l}) reflection for the same films. 
}
\label{figDWexp}
\end{figure}

Regarding the magnetic characterization, the samples were measured in a longitudinal MOKE setup equipped with an in-plane rotation stage inside a cryostat [schematic in Fig.~2 a)]. 
This configuration allows temperature sweeps to be performed with or without an applied magnetic field while maintaining full in-plane rotational control of the sample \cite{cunado2017}.
Prior to the angular measurements, the films were field-cooled from room temperature down to 50\,K under an in-plane magnetic field of 15\,mT. 
Once the temperature was stabilized, hysteresis loops were recorded while rotating the sample through 360$^\circ$. 
Representative loops of 7\,nm sample measured at $-90^\circ$, $0^\circ$, and $+90^\circ$ are shown in Figures~2 b--d), corresponding respectively to the magnetization easy axis, hard axis, and the complementary easy axis of the FM layer. Along the easy axis, the hysteresis loop exhibits a clear horizontal shift due to exchange bias \cite{jimenez2009, perna2015}, with a magnitude of approximately 0.35\,mT. The abrupt irreversible transitions reveal a magnetization reversal pathways dominated by nucleation and propagation of magnetic domains \cite{pernaNJP2010, pernaJAP2011}.
At the hard axis, no measurable bias is observed, as expected for a uniaxial anisotropy \cite{jimenez2009, perna2015}. Fully reversible magnetization transitions are associated to coherent rotation of magnetic domains \cite{pernaNJP2010, pernaJAP2011}.

A key signature of genuine exchange bias is the reversal of the loop shift upon rotating the sample by 180$^\circ$ \cite{jimenez2009}. 
This inversion is clearly observed when comparing the $-90^\circ$ and $+90^\circ$ loops, demonstrating that the bias originates from the sample and not from spurious remanence of the electromagnet coils.
The full angular evolution of the exchange-bias field $\mu_0 H_{\mathrm{EB}}$ extracted from all measured orientations is shown in Figure~2 e). The sinusoidal-like dependence and the sign reversal every 180$^\circ$ confirm the intrinsic nature of the exchange bias in these ultra-thin LSMO films.

\begin{figure}[t]
\centering
\includegraphics[width=\columnwidth]{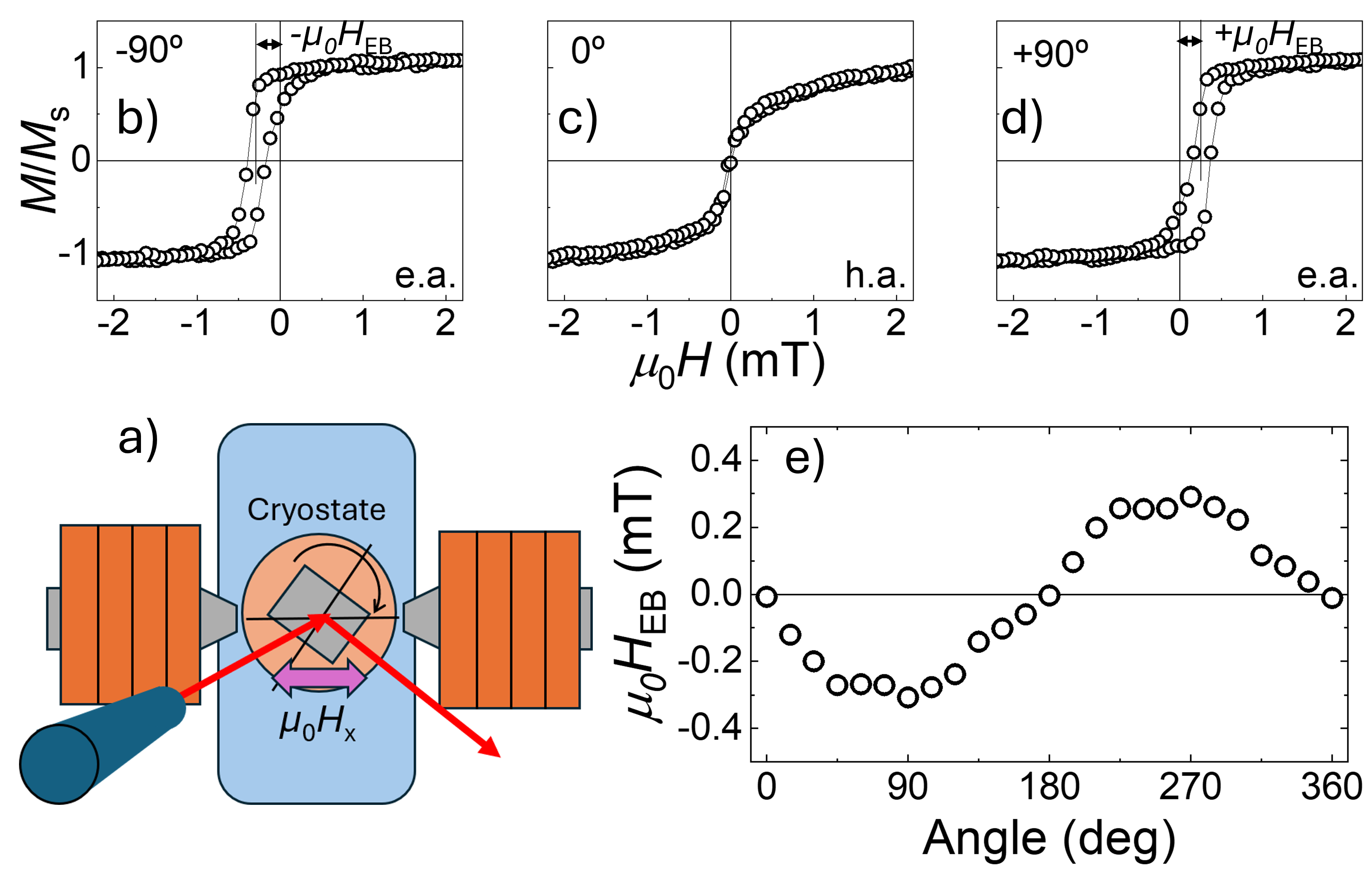}
\caption{
a) Schematic of the longitudinal MOKE setup used for angular measurements. 
b–d) Normalized hysteresis loops $M/M_{\mathrm{s}}$ measured at selected angles ($-90^\circ$, $0^\circ$, and $+90^\circ$). 
A clear exchange-bias shift is observed along the easy axis, while the hard-axis loop shows no measurable bias. 
e) Angular dependence of the exchange-bias field $\mu_{0}H_{\mathrm{EB}}$, exhibiting a sinusoidal-like evolution and a sign inversion upon rotation, demonstrating that the EB is intrinsic and reverses consistently with sample orientation.
}
\label{fig:MOKE_EB_angular}
\end{figure}

We now analyze the temperature dependence of the exchange bias in the three LSMO films. As summarized in Figure~3 a-c) that show representative hysteresis loops for the 3\,nm film at 65, 150, and 200\,K (black circles). 
A clear reduction of the coercive field is observed as the temperature approaches the Curie temperature, 
$T_{\mathrm{C}} \approx 160$\,K, extracted from the $M(T)$ curve in Figure~3 j). 
At 200\,K the film is already in the paramagnetic state, and therefore no hysteresis loop is present. 
Despite the strong temperature dependence of the coercivity, the exchange-bias amplitude remains nearly constant at the temperatures where the film is in ferromagnetic state.
Panels~3(d)–3(f) display the loops of the 7\,nm film (red symbols) at 110, 180, and 250\,K. 
The Curie temperature of this sample, obtained from Figure~3 k), is $T_{\mathrm{C}} \approx 250$\,K. 
The coercive field again decreases as $T$ approaches $T_{\mathrm{C}}$, while the exchange-bias field remains of similar magnitude across the entire temperature range.
Finally, the 17\,nm film is shown in Figures~3 g)–3 i) (blue symbols) at 100, 180, and 270\,K. 
Step-like features (potentially originating from the coexistence of different underlying phases) appear in the hysteresis loops, more prominently than in the thinner films. However, the exchange-bias shift is still clearly resolved and remains approximately constant or even slightly increases as the temperature approaches $T_{\mathrm{C}} \approx 310$\,K, as inferred from Figure~3 l).
The extracted $\mu_0 H_{\mathrm{EB}}$ for all three thicknesses are plotted as a function of temperature in Figure~3 m), using the same color code as the loops. 
Although the data show some scatter, the magnitude of $\mu_0 H_{\mathrm{EB}}$ is essentially thickness-independent and only weakly dependent on temperature. This observation is consistent with a non‑interfacial origin, given the lack of thickness dependence.
The field-cooled (FC) and zero-field-cooled (ZFC) $M(T)$ curves for the three films, shown in Figures~3 j)–3 l), reveal a pronounced divergence between FC (red) and ZFC (black) branches. 
This effect is strongest in the 7\,nm film, where the ZFC magnetization even reverses sign at low temperature, and is also clearly visible in the 17\,nm film. 
All samples exhibit a characteristic feature around the Néel temperature $T_{\mathrm{N}} \approx 100$\,K, consistent with the presence of antiferromagnetic-like regions. 
The strong FC–ZFC irreversibility and the sign reversal of the ZFC magnetization provide compelling evidence for the coexistence of competing magnetic phases, which likely underpins the emergence of exchange bias in these ultrathin LSMO films.

\begin{figure*}[t]
\centering
\includegraphics[width=\textwidth]{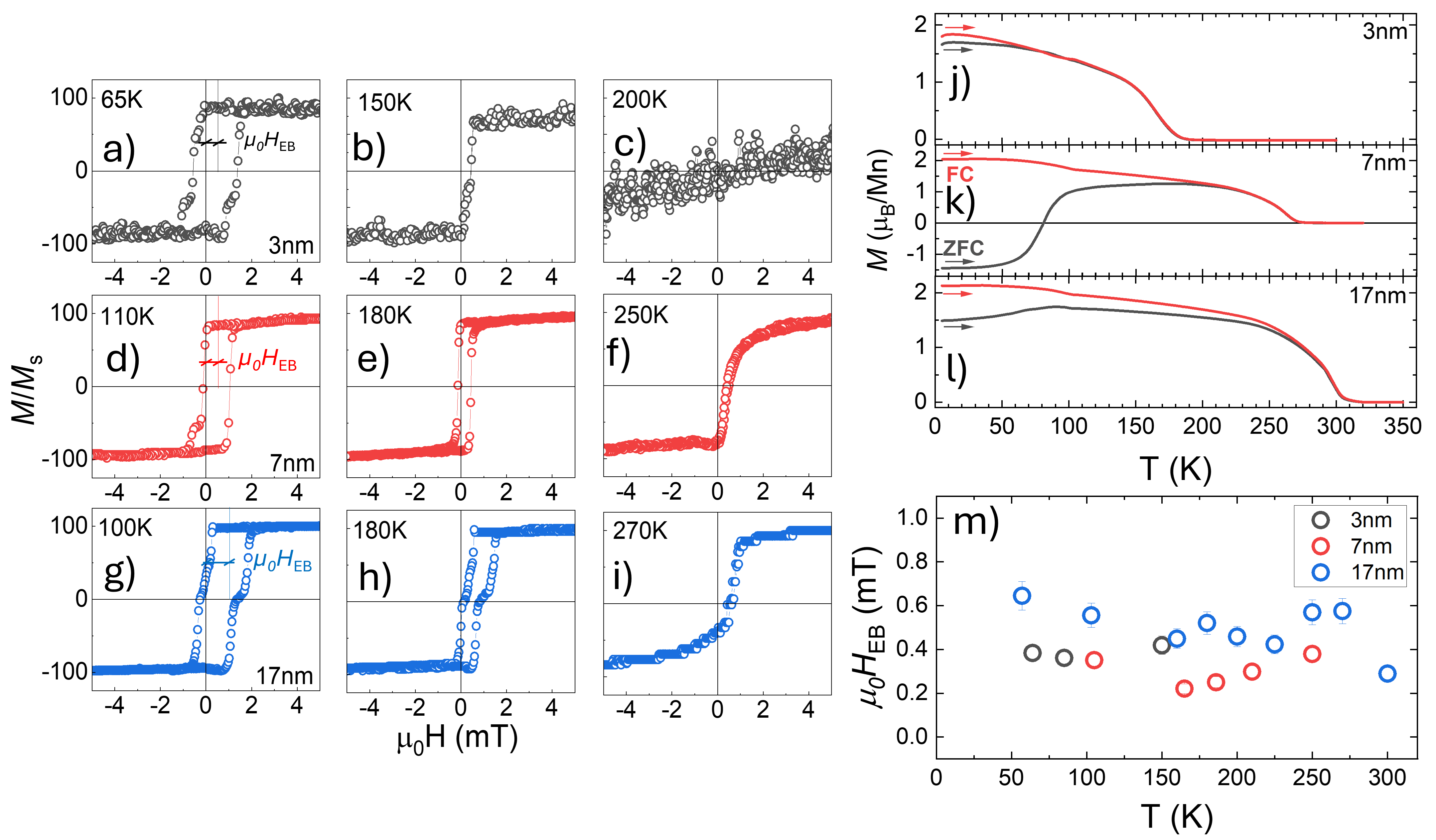}
\caption{
 a–i) Temperature-dependent hysteresis loops for 3, 7, and 17 nm LSMO films. Coercivity decreases as $T$ approaches $T_C$, while the exchange-bias shift remains visible across the ferromagnetic regime. Step-like features appear more prominently in the 17 nm film. j–l) ZFC–FC magnetization curves showing strong irreversibility in all films and a low-temperature magnetization reversal in the 7 nm sample. m) Temperature evolution of $H_{\mathrm{EB}}$ for all thicknesses, revealing a nearly thickness-independent bias with weak temperature dependence.
}
\label{fig:LSMO_magnetism_temperature}
\end{figure*}

\section{Discussion}

To elucidate the micromagnetic origin of the exchange bias in single-layer LSMO films, we performed a detailed chemical analysis by XPS. 
Figure~4 displays the Mn\,2$p$ and Sr\,3$d$ core-level spectra for the three film thicknesses. 
In stoichiometric La$_{0.67}$Sr$_{0.33}$MnO$_3$, charge neutrality dictates that the substitution of La$^{3+}$ by Sr$^{2+}$ must be compensated by the oxidation of Mn ions. Consequently, the Mn valence is fixed by the Sr content: one third of the Mn ions adopt the Mn$^{4+}$ state, while the remaining two thirds remain as Mn$^{3+}$. This corresponds to an average Mn valence of 3.33 and a well-defined Mn$^{3+}$/Mn$^{4+}$ ratio of 2:1. Any deviation from this proportion therefore provides a direct signature of oxygen non-stoichiometry. In our films, the XPS spectra reveal a substantial Mn$^{2+}$ component of approximately 20\%, together with a reduced Mn$^{3+}$/Mn$^{4+}$ ratio, indicating that the Mn valence is significantly lower than the expected 3.33. 
The Mn$^{3+}$ and Mn$^{4+}$ components are also resolved, and their relative ratio remains nearly constant across the three thicknesses. 
This thickness-independent chemical landscape demonstrates that the degree of reduction does not depend on film thickness and that the oxygen-deficient regions are distributed throughout the film volume rather than confined to the interfaces.  

The Sr\,3$d$ spectra, shown in Figures~4 d)–4 f), display the expected Sr$^{2+}$ doublet together with a small SrO-related contribution. In particular, the SrO components are significantly weaker in the 17~nm film than in the thinner ones, indicating that Sr segregation or SrO surface formation is strongly suppressed at larger thickness. This supports the view that the dominant non-stoichiometric effects originate from the Mn--O sublattice rather than from Sr chemistry.

The coexistence of a substantial Mn$^{2+}$ fraction with a largely thickness-independent Mn$^{3+}$/Mn$^{4+}$ ratio reveals a phase coexistent landscape composed of reduced inclusions embedded within a more oxidized LSMO-like matrix. 
Such chemical inhomogeneity \cite{capua2006} naturally produces a mixed magnetic state: Mn$^{2+}$-rich regions behave as antiferromagnetic clusters, while the Mn$^{3+}$/Mn$^{4+}$ matrix remains ferromagnetic. 
The magnetic response of each film reflects the connectivity of this ferromagnetic matrix. 
At 3\,nm the FM network could not form a continuous connected pathway, and the AFM component dominates, yielding weak magnetization and moderate ZFC--FC splitting  below $T_\mathrm{N}$.
The 7\,nm film corresponds to an intermediate regime in which the FM matrix is continuous but still strongly intermixed with Mn$^{2+}$-rich AFM islands. This configuration maximizes AFM$|$FM coupling and produces strong competition between the two phases, as evidenced by the large ZFC--FC divergence and the negative magnetization at low temperature, which is driven by the unidirectional exchange anisotropy at the phase boundaries. The energy associated with the pinned FM$|$AFM interfaces creates a local energy barrier that prevents the FM spins from rotating along the weak measuring field, stabilizing an inverted magnetic state at low temperatures~\cite{saghayezhian2019a, mottaghi2018a}.
At 17\,nm the FM matrix is fully developed and effectively masks the AFM clusters, producing a more bulk-like response in which the reduced regions cannot reverse the magnetization under small fields.

The Mn valence extracted from the XPS $\text{Mn}^{2+}$/$\text{Mn}^{3+}$/$\text{Mn}^{4+}$ ratios yields an average oxidation state of $v_{\mathrm{Mn}} \approx 2.69$--$2.75$. Satisfying global charge neutrality in $\text{La}_{0.67}\text{Sr}_{0.33}\text{MnO}_{3-\delta}$ requires balancing the cationic and anionic charges according to:
\[
0.67(+3) + 0.33(+2) + v_{\mathrm{Mn}} + (3-\delta)(-2) = 0
\]
which directly leads to:
\[
\delta = \frac{3.33 - v_{\mathrm{Mn}}}{2} \approx 0.29\text{--}0.32
\]
Such an intermediate average valence confirms the phase coexistence landscape~\cite{dagotto2001a}. In this scenario, $\text{Mn}^{2+}$-rich reduced island-like coexist with a pristine ferromagnetic matrix where the $\text{Mn}^{3+}$/$\text{Mn}^{4+}$ ratio remains close to unity ($v_{\mathrm{Mn}} \sim 3.3$). This mixed-phase behavior is reminiscent of oxygen-deficient, brownmillerite-derived manganites~\cite{sakai2021, gorzkowski2026}. In these systems, severe structural distortions and the resulting local $\text{Mn}^{2+}$/$\text{Mn}^{3+}$ mixtures cooperate to suppress double-exchange conduction, thereby stabilizing G- or C-type antiferromagnetic order (or, depending on the local geometry and vacancy arrangement, even spin‑ice‑like frustrated configurations) within the reduced insulating islands~\cite{chen2019, peng2020, ali2007}, which subsequently act as the pinning phase responsible for the observed exchange bias effect~\cite{wisniewski2017}. 
The resulting phase competition accounts for the exchange bias, the strong ZFC–FC irreversibility, and the magnetization inversion observed at intermediate thickness.

\begin{figure}[t]
\centering
\includegraphics[width=\columnwidth]{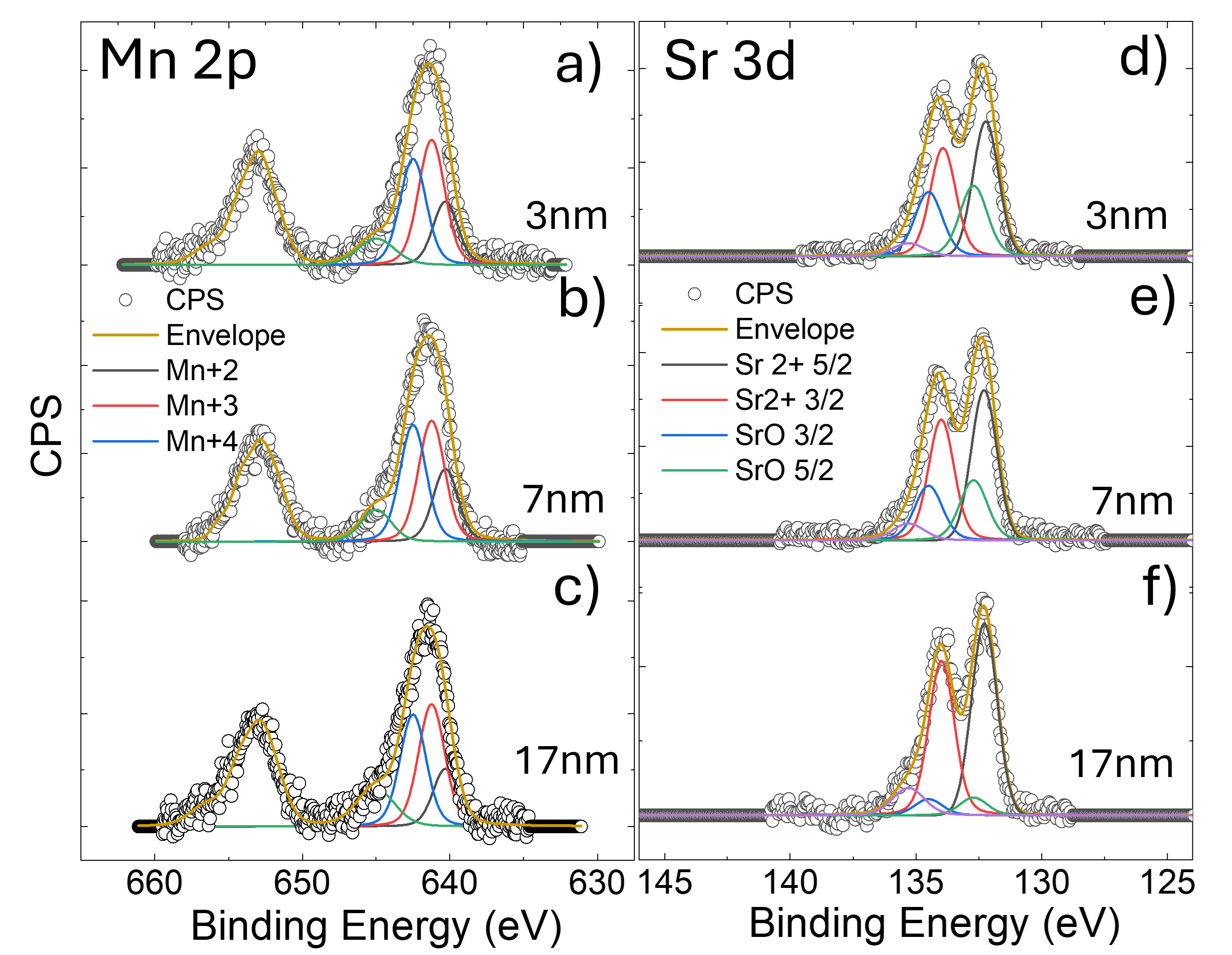}
\caption{
a–c) Mn\,2$p$ XPS spectra for La$_{0.67}$Sr$_{0.33}$MnO$_3$ films with thicknesses of 
3 nm, 7 nm, and 17 nm. Open circles correspond to the experimental counts (CPS), while solid lines represent the fitted envelope (yellow) and individual chemical components: Mn$^{2+}$ (black), Mn$^{3+}$ (red), Mn$^{4+}$ (blue), and the shake-up satellite. 
All films exhibit a substantial Mn$^{2+}$ contribution, indicating oxygen-deficient reduced regions coexisting with the Mn$^{3+}$/Mn$^{4+}$ matrix.
d–f) Sr\,3$d$ XPS spectra for the same films, showing the fitted envelope together with the Sr$^{2+}$ 3$d_{5/2}$ and 3$d_{3/2}$ doublet, SrO-related components, and the shake-up feature. 
}
\label{fig:LSMO_XPS}
\end{figure}

The small amplitude of exchange-bias field showed in Figure~3 m) allow us to estimate the effective amount of AFM phase that is magnetically coupled to the FM LSMO. For this purpose, we use the standard interfacial exchange expression
\begin{equation}
\mu_0 H_\mathrm{EB} = \frac{f_\mathrm{AFM} J_\mathrm{int}^0}{M_\mathrm{FM} t_\mathrm{FM}}
\end{equation}

where $f_\mathrm{AFM}$ denotes the effective AFM area fraction contributing to the exchange bias, $J_\mathrm{int}^0$ is the interfacial exchange energy density, $M_\mathrm{FM}$ is the magnetization of the LSMO layer, and $t_\mathrm{FM}$ its thickness. Using the measured $H_\mathrm{EB}$ values together with the magnetic moments per Mn extracted at $T_\mathrm{C}$, we convert the moments (0.452, 0.444, and 0.627~$\mu_\mathrm{B}$/Mn for 3, 7, and 17~nm) into volumetric magnetizations by assuming a pseudocubic lattice parameter $a\approx 3.9$~\AA. Inserting these values into the expression above and adopting a representative $J_\mathrm{int}^0 = 0.5$~mJ/m$^2$ \cite{nogues1999, ziese2010a}, we obtain effective AFM fractions in the range $f_\mathrm{AFM}\sim 10^{-4}$–$10^{-3}$ for the three thicknesses. Such sub-percent coverage implies that only a very small fraction of the interface needs to host AFM regions to generate the observed bias. This picture is fully consistent with the chemical analysis: although the Mn$^{2+}$ fraction is of the order of tens of percent, only a minor subset of these locally oxygen-deficient regions must develop AFM correlations and couple to the LSMO. 
The exchange bias therefore supports a scenario of nanoscale, non-uniform AFM islands rather than a continuous AFM layer.

Additional insight into the phase coexistence is provided by the transport measurements of the 3\,nm film, shown in Fig.~5. The $I$--$V$ curves display a linear behavior up to $\sim 20$\,V, followed by resistive switching and a negative differential resistance regime above $\sim 40$\,V, likely driven by local Joule heating and volatile changes that make the material more insulating before the voltage is reduced coming back to its initial state. The inset in Fig.~5 a) shows in detail the presence of step-like features. The temperature dependence of the step-like feature further supports this picture: it is most pronounced at low temperature, when the LSMO matrix is highly metallic and the contrast with the insulating clusters is maximal, and it disappears as the temperature approaches $T_\mathrm{C}$, where the film becomes nearly insulating and the switching vanishes. Although transport cannot unambiguously determine the magnetic nature of these regions, the hysteresis and discrete conduction steps point to an inhomogeneous landscape where reduced, insulating AFM-like inclusions interrupt the ferromagnetic network.

\begin{figure}[t]
\centering
\includegraphics[width=\columnwidth]{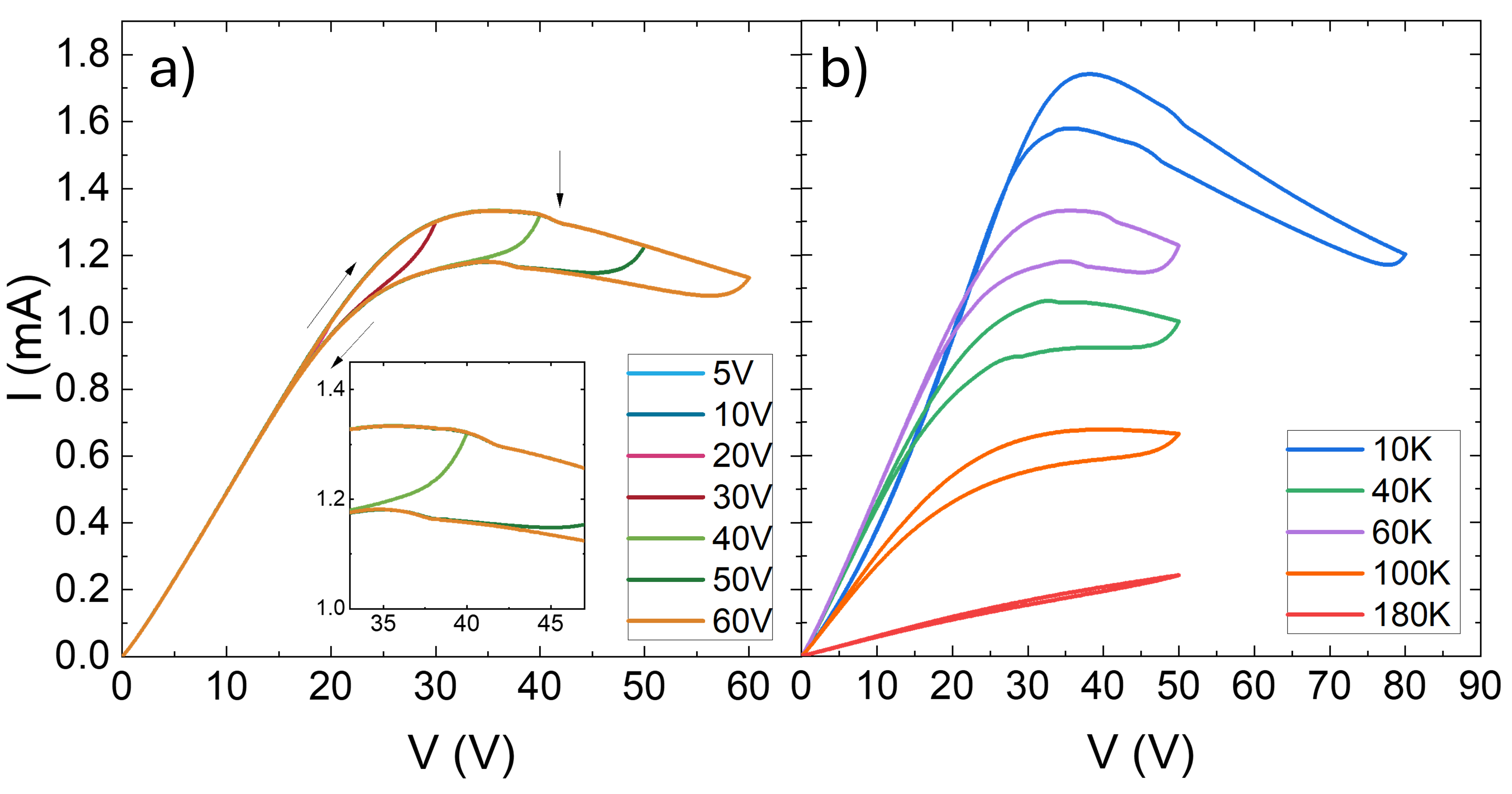}
\caption{
 a) Nonlinear and hysteretic I–V characteristics of the 3 nm film, with a step-like (magnified in the inset) increase in current at high bias consistent with transport through reduced insulating regions. b) Temperature-dependent I–V curves showing enhanced nonlinearity and hysteresis at low temperature, supporting a granular landscape where oxygen-deficient domains interrupt the conductive ferromagnetic matrix.
}
\label{fig:LSMO_IV}
\end{figure}

\section{Conclusion}

Ultra-thin La$_{0.67}$Sr$_{0.33}$MnO$_3$ films exhibit a robust and thickness-independent exchange bias despite being nominally single-layer ferromagnets. A comprehensive combination of structural, magnetic, chemical, and transport measurements demonstrates that this behavior originates from an intrinsic phase-coexistence state driven by oxygen deficiency. XPS reveals a thickness-independent Mn$^{2+}$ fraction, evidencing reduced regions embedded within a Mn$^{3+}$/Mn$^{4+}$ ferromagnetic matrix. These reduced regions behave as antiferromagnetic clusters, chemically analogous to brownmillerite-like inclusions, and persist across all thicknesses.

The magnetic response evolves with the connectivity of the ferromagnetic matrix. At 3\,nm the FM network is incomplete and the reduced AFM component dominates, whereas at 17\,nm the FM matrix is fully developed and largely screens the AFM clusters. The 7\,nm film represents an intermediate regime in which the FM matrix is continuous but still strongly intermixed with Mn$^{2+}$-rich AFM islands, maximizing AFM--FM competition. This regime produces the strongest FC--ZFC irreversibility, a magnetization sign inversion, and the clearest manifestation of exchange bias. Angular MOKE measurements further confirm the intrinsic nature of the effect through the characteristic 180$^\circ$ inversion of the loop shift.

Transport measurements reinforce this picture by revealing nonlinear, thermally activated conduction and discrete step-like features, consistent with insulating reduced regions interrupting the conducting FM network, which furthermore enable non-volatile resistive switching behavior. Although transport does not directly probe the magnetic structure, it corroborates the inhomogeneous chemical landscape inferred from XPS and magnetometry.

Altogether, these results show that exchange bias in ultrathin LSMO does not require mandatorily engineered interfaces but may emerge from self-organized AFM clusters generated by oxygen deficiency within a single film. This mechanism highlights the extreme sensitivity of correlated oxides to subtle variations in oxygen stoichiometry and demonstrates that robust exchange-bias phenomena typically associated to interface magnetism arise spontaneously even in single layers. These findings open new routes for controlling exchange bias and phase competition in manganites through growth conditions, oxygen engineering, and dimensionality.

\section{Supplementary Material}
See the supplementary material for additional characterization, and supporting figures complementing the results presented in the main text.

\begin{acknowledgments}
This research was supported by the Spanish AEI/MICINN through projects MERANET “GREENMEM” PCI2025-163177, "HYPERFAN" PID2024-157112OB-C53 and “PHANTOMA” PID2024-157080OA-C32. The authors also acknowledge support from the Comunidad de Madrid through projects TEC-2024/TEC-380 “Mag4TIC” and Atracción de Talento 2022-T1/IND-23901. Work at UCM supported by Agencia Estatal de Investigación through PID2023-148884OB-I00 and PID2024-155385NB-C31
\end{acknowledgments}
\bibliography{bibliography.bib}

@article{Zener1951,
  author    = {Zener, Clarence},
  title     = {Interaction Between the d-Shells in the Transition Metals},
  journal   = {Phys. Rev.},
  volume    = {81},
  pages     = {440--444},
  year      = {1951},
  doi       = {10.1103/PhysRev.81.440}
}

@article{Anderson1955,
  author    = {Anderson, P. W. and Hasegawa, H.},
  title     = {Considerations on Double Exchange},
  journal   = {Phys. Rev.},
  volume    = {100},
  pages     = {675--681},
  year      = {1955},
  doi       = {10.1103/PhysRev.100.675}
}

@article{Millis1995,
  author    = {Millis, A. J. and Littlewood, P. B. and Shraiman, B. I.},
  title     = {Double exchange alone does not explain the resistivity of La$_{1-x}$Sr$_x$MnO$_3$},
  journal   = {Phys. Rev. Lett.},
  volume    = {74},
  pages     = {5144--5147},
  year      = {1995},
  doi       = {10.1103/PhysRevLett.74.5144}
}

@article{Coey1999,
  author    = {Coey, J. M. D. and Viret, M. and von Moln{\'a}r, S.},
  title     = {Mixed-Valence Manganites},
  journal   = {Adv. Phys.},
  volume    = {48},
  pages     = {167--293},
  year      = {1999},
  doi       = {10.1080/000187399243455}
}

@article{Tokura2000,
  author    = {Tokura, Yoshinori and Nagaosa, Naoto},
  title     = {Orbital Physics in Transition-Metal Oxides},
  journal   = {Science},
  volume    = {288},
  pages     = {462--468},
  year      = {2000},
  doi       = {10.1126/science.288.5465.462}
}

@article{Dagotto2001,
  author    = {Dagotto, Elbio and Hotta, Takashi and Moreo, Adriana},
  title     = {Colossal Magnetoresistant Materials: The Key Role of Phase Separation},
  journal   = {Phys. Rep.},
  volume    = {344},
  pages     = {1--153},
  year      = {2001},
  doi       = {10.1016/S0370-1573(00)00121-6}
}

@article{trappen2019,
  title={Effect of oxygen stoichiometry on the magnetization profiles and negative magnetization in LSMO thin films},
  author={Trappen, Robbyn and Grutter, Alexander J and Huang, Chih-Yeh and Penn, Aubrey and Mottaghi, Navid and Yousefi, Saeed and Haertter, Allison and Kumari, Shalini and LeBeau, James and Kirby, Brian J and others},
  journal={Journal of Applied Physics},
  volume={126},
  number={10},
  year={2019},
  publisher={AIP Publishing}
}

@article{salev2021,
  title={Transverse barrier formation by electrical triggering of a metal-to-insulator transition},
  author={Salev, Pavel and Fratino, Lorenzo and Sasaki, Dayne and Berkoun, Rani and Del Valle, Javier and Kalcheim, Yoav and Takamura, Yayoi and Rozenberg, Marcelo and Schuller, Ivan K},
  journal={Nature communications},
  volume={12},
  number={1},
  pages={5499},
  year={2021},
  publisher={Nature Publishing Group UK London}
}

@article{chen2019,
  title={Thickness-driven first-order phase transitions in manganite ultrathin films},
  author={Chen, Hongyan and Yu, Yang and Wang, Zhe and Bai, Yu and Lin, Hanxuan and Li, Xiaolong and Liu, Hao and Miao, Tian and Kou, Yunfang and Zhang, Yongsheng and others},
  journal={Physical Review B},
  volume={99},
  number={21},
  pages={214419},
  year={2019},
  publisher={APS}
}

@article{tebano2006,
  title={Strain-induced phase separation in La 0.7 Sr 0.3 Mn O 3 thin films},
  author={Tebano, A and Aruta, C and Medaglia, PG and Tozzi, F and Balestrino, G and Sidorenko, AA and Allodi, Giuseppe and De Renzi, Roberto and Ghiringhelli, G and Dallera, Claudia and others},
  journal={Physical Review B—Condensed Matter and Materials Physics},
  volume={74},
  number={24},
  pages={245116},
  year={2006},
  publisher={APS}
}

@article{lopez2023,
  title={Electrolyte Gated Synaptic Transistor based on an Ultra-Thin Film of La0. 7Sr0. 3MnO3},
  author={L{\'o}pez, Alejandro and Tornos, Javier and Peralta, Andrea and Barbero, Isabel and Fernandez-Canizares, Francisco and Sanchez-Santolino, Gabriel and Varela, Mar{\'\i}a and Rivera, Alberto and Camarero, Julio and Le{\'o}n, Carlos and others},
  journal={Advanced Electronic Materials},
  volume={9},
  number={7},
  pages={2300007},
  year={2023},
  publisher={Wiley Online Library}
}

@article{Perna2017,
author = {Perna, Paolo and Maccariello, Davide and Ajejas, Fernando and Guerrero, Ruben and Méchin, Laurence and Flament, Stephane and Santamaria, Jacobo and Miranda, Rodolfo and Camarero, Julio},
title = {Engineering Large Anisotropic Magnetoresistance in La0.7Sr0.3MnO3 Films at Room Temperature},
journal = {Advanced Functional Materials},
volume = {27},
number = {26},
pages = {1700664},
doi = {https://doi.org/10.1002/adfm.201700664},
year = {2017},
}

@article{Perna2012,
  title = {Magnetization reversal signatures in the magnetoresistance of magnetic multilayers},
  author = {Perna, P. and Rodrigo, C. and Mu\~noz, M. and Prieto, J. L. and Bollero, A. and Maccariello, D. and Cu\~nado, J. L. F. and Romera, M. and Akerman, J. and Jim\'enez, E. and Mikuszeit, N. and Cros, V. and Camarero, J. and Miranda, R.},
  journal = {Phys. Rev. B},
  volume = {86},
  issue = {2},
  pages = {024421},
  numpages = {4},
  year = {2012},
  month = {Jul},
  publisher = {American Physical Society},
  doi = {10.1103/PhysRevB.86.024421},
  url = {https://link.aps.org/doi/10.1103/PhysRevB.86.024421}
}

@article{pernaNJP2010,
doi = {10.1088/1367-2630/12/10/103033},
url = {https://doi.org/10.1088/1367-2630/12/10/103033},
year = {2010},
month = {oct},
publisher = {},
volume = {12},
number = {10},
pages = {103033},
author = {Perna, P and Méchin, L and Saïb, M and Camarero, J and Flament, S},
title = {Imaging the magnetization reversal of step-induced uniaxial magnetic anisotropy in vicinal epitaxial La0.7Sr0.3MnO3 films},
journal = {New Journal of Physics}
}

@article{pernaJAP2011,
    author = {Perna, P. and Rodrigo, C. and Jiménez, E. and Teran, F. J. and Mikuszeit, N. and Méchin, L. and Camarero, J. and Miranda, R.},
    title = {Tailoring magnetic anisotropy in epitaxial half metallic La0.7Sr0.3MnO3 thin films},
    journal = {Journal of Applied Physics},
    volume = {110},
    number = {1},
    pages = {013919},
    year = {2011},
    month = {07},
    issn = {0021-8979},
    doi = {10.1063/1.3605542},
    url = {https://doi.org/10.1063/1.3605542},
    eprint = {https://pubs.aip.org/aip/jap/article-pdf/doi/10.1063/1.3605542/13241920/013919_1_online.pdf},
}

@article{jimenez2009,
    author = {Jiménez, E. and Camarero, J. and Sort, J. and Nogués, J. and Hoffmann, A. and Teran, F. J. and Perna, P. and García-Martín, J. M. and Dieny, B. and Miranda, R.},
    title = {Highly asymmetric magnetic behavior in exchange biased systems induced by noncollinear field cooling},
    journal = {Applied Physics Letters},
    volume = {95},
    number = {12},
    pages = {122508},
    year = {2009},
    month = {09},
    issn = {0003-6951},
    doi = {10.1063/1.3236768},
    url = {https://doi.org/10.1063/1.3236768},
    eprint = {https://pubs.aip.org/aip/apl/article-pdf/doi/10.1063/1.3236768/14419204/122508_1_online.pdf},
}

@article{chaluvadi2020,
  title={Epitaxial strain and thickness dependent structural, electrical and magnetic properties of La0. 67Sr0. 33MnO3 films},
  author={Chaluvadi, Sandeep Kumar and Ajejas, Fernando and Orgiani, Pasquale and Lebargy, Sylvain and Minj, Albert and Flament, St{\'e}phane and Camarero, Julio and Perna, Paolo and M{\'e}chin, Laurence},
  journal={Journal of Physics D: Applied Physics},
  volume={53},
  number={37},
  pages={375005},
  year={2020},
  publisher={IOP Publishing}
}

@article{navarro2023,
  title={Light-induced decoupling of electronic and magnetic properties in manganites},
  author={Navarro, Henry and Basaran, Ali C and Ajejas, Fernando and Fratino, Lorenzo and Bag, Soumen and Wang, Tianxing D and Qiu, Erbin and Rouco, Victor and Tenreiro, Isabel and Torres, Felipe and others},
  journal={Physical Review Applied},
  volume={19},
  number={4},
  pages={044077},
  year={2023},
  publisher={APS}
}

@article{nogues1999,
  title={Exchange bias},
  author={Nogu{\'e}s, Josep and Schuller, Ivan K},
  journal={Journal of Magnetism and Magnetic Materials},
  volume={192},
  number={2},
  pages={203--232},
  year={1999},
  publisher={Elsevier}
}

@article{niebieskikwiat2005,
  title={Intrinsic interface exchange coupling of ferromagnetic nanodomains in a charge ordered manganite},
  author={Niebieskikwiat, D and Salamon, MB},
  journal={Physical Review B—Condensed Matter and Materials Physics},
  volume={72},
  number={17},
  pages={174422},
  year={2005},
  publisher={APS}
}

@article{dong2008,
  title={Magnetism, conductivity, and orbital order in (LaMnO 3) 2 n/(SrMnO 3) n superlattices},
  author={Dong, Shuai and Yu, Rong and Yunoki, Seiji and Alvarez, Gonzalo and Liu, J-M and Dagotto, Elbio},
  journal={Physical Review B—Condensed Matter and Materials Physics},
  volume={78},
  number={20},
  pages={201102},
  year={2008},
  publisher={APS}
}

@article{Zhang2021,
  author    = {Zhang, Ji and Yang, Jack and Causer, Grace L. and Shi, Junjie and Klose, Frank et al.},
  title     = {Realization of exchange bias control via interfacial frustration in magnetic complex oxide heterostructures},
  journal   = {Phys. Rev. B},
  volume    = {104},
  pages     = {174444},
  year      = {2021},
  doi       = {10.1103/PhysRevB.104.174444}
}

@article{wu2020,
  title={Strain-induced exchange bias transition at La0. 7Sr0. 3MnO3/NiO interface},
  author={Wu, Yingjie and Wang, Zhanjie and Liang, Yongmei and Zhang, Zhidong},
  journal={Journal of Alloys and Compounds},
  volume={819},
  pages={153022},
  year={2020},
  publisher={Elsevier}
}

@article{Zhang2024,
  author    = {Zhang, Jun and Su, Tiancong and Ma, Jianchun},
  title     = {Strain-induced robust exchange bias in epitaxial La$_{0.7}$Sr$_{0.3}$MnO$_3$/LaFeO$_3$ bilayers},
  journal   = {Molecules},
  volume    = {29},
  number    = {14},
  pages     = {3244},
  year      = {2024},
  doi       = {10.3390/molecules29143244}
}

@article{Ziese2010,
  author    = {Ziese, M. and Vrejoiu, I. and Hesse, D.},
  title     = {Exchange bias in La$_{0.7}$Sr$_{0.3}$MnO$_3$/SrRuO$_3$ heterostructures},
  journal   = {Phys. Rev. B},
  volume    = {81},
  pages     = {184418},
  year      = {2010},
  doi       = {10.1103/PhysRevB.81.184418}
}

@article{Bibes2008,
  author    = {Bibes, Manuel and Barth{\'e}l{\'e}my, Agn{\`e}s},
  title     = {Oxide spintronics},
  journal   = {Nat. Mater.},
  volume    = {7},
  pages     = {425--428},
  year      = {2008},
  doi       = {10.1038/nmat2189}
}

@article{perna2015,
  title={Interfacial exchange-coupling induced chiral symmetry breaking of spin-orbit effects},
  author={Perna, Paolo and Ajejas, F and Maccariello, D and Fernandez Cu{\~n}ado, JL and Guerrero, R and Ni{\~n}o, MA and Bollero, Alberto and Miranda, Rodolfo and Camarero, Julio},
  journal={Physical Review B},
  volume={92},
  number={22},
  pages={220422},
  year={2015},
  publisher={APS}
}

@article{chaluvadi2022,
  title={Electronic Properties of Fully Strained La1--x Sr x MnO3 Thin Films Grown by Molecular Beam Epitaxy},
  author={Chaluvadi, Sandeep Kumar and Polewczyk, Vincent and Petrov, Aleksandr Yu and Vinai, Giovanni and Braglia, Luca and Diez, Jose Manuel and Pierron, Victor and Perna, Paolo and Mechin, Laurence and Torelli, Piero and others},
  journal={ACS omega},
  volume={7},
  number={17},
  pages={14571--14578},
  year={2022},
  publisher={ACS Publications}
}

@article{ajejas2023,
  title={Current-Driven Switching of N{\'e}el Vector of an Antiferromagnetic Insulator Thin Film},
  author={Ajejas, Fernando and Torres, Felipe and Basaran, Ali C and Salev, Pavel and Schuller, Ivan K},
  journal={Advanced Electronic Materials},
  volume={9},
  number={11},
  pages={2300385},
  year={2023},
  publisher={Wiley Online Library}
}

@article{cunado2015,
  title={Note: Vectorial-magneto optical Kerr effect technique combined with variable temperature and full angular range all in a single setup},
  author={Cu{\~n}ado, Jose Luis F and Pedrosa, Javier and Ajejas, Fernando and Bollero, Alberto and Perna, Paolo and Teran, Francisco J and Miranda, Rodolfo and Camarero, Julio},
  journal={Review of Scientific Instruments},
  volume={86},
  number={4},
  year={2015},
  publisher={AIP Publishing}
}

@article{cunado2017,
  title={Direct observation of temperature-driven magnetic symmetry transitions by vectorial resolved MOKE magnetometry},
  author={Luis F Cu{\~n}ado, Jose and Pedrosa, Javier and Ajejas, Fernando and Perna, Paolo and Miranda, Rodolfo and Camarero, Julio},
  journal={Journal of Physics: Condensed Matter},
  volume={29},
  number={40},
  pages={405805},
  year={2017},
  publisher={IOP Publishing}
}

@article{dagotto2001a,
  author    = {Dagotto, Elbio and Hotta, Takashi and Moreo, Adriana},
  title     = {Phase separation in manganites and other strongly correlated electron systems},
  journal   = {Physics Reports},
  volume    = {344},
  number    = {1-3},
  pages     = {1--153},
  year      = {2001},
  publisher = {Elsevier}
}

@article{sakai2021,
  author    = {Sakai, E. and Tamamitsu, M. and Yoshimatsu, K. and Oshima, M. and Kumigashira, H.},
  title     = {Gradual transition from ferromagnetic metal to G-type antiferromagnetic insulator in ultra-reduced $\mathrm{La}_{0.7}\mathrm{Sr}_{0.3}\mathrm{MnO}_{3-\delta}$ thin films},
  journal   = {Physical Review Materials},
  volume    = {5},
  number    = {8},
  pages     = {084406},
  year      = {2021},
  publisher = {APS}
}

@article{gorzkowski2026,
  author    = {Gorzkowski, M. and Tanaka, H. and Martinez, L.},
  title     = {Oxygen vacancy clustering and G-type antiferromagnetism in epitaxial manganite thin films},
  journal   = {Physical Review B},
  volume    = {113},
  number    = {4},
  pages     = {045112},
  year      = {2026},
  publisher = {APS}
}

@article{peng2020,
  author    = {Peng, J. and Song, C. and Li, F. and Wang, B. and Pan, F.},
  title     = {Origin of insulating weak ferromagnetic phase and G-type antiferromagnetism in manganite interfaces},
  journal   = {AIP Advances},
  volume    = {10},
  number    = {5},
  pages     = {055114},
  year      = {2020},
  publisher = {AIP Publishing}
}

@article{ziese2010a,
  author    = {Ziese, M. and Vrejoiu, I. and Hesse, D.},
  title     = {Exchange bias in $\mathrm{La}_{0.7}\mathrm{Sr}_{0.3}\mathrm{MnO}_3$/$\mathrm{SrRuO}_3$ superlattices},
  journal   = {Physical Review B},
  volume    = {81},
  number    = {18},
  pages     = {184418},
  year      = {2010},
  publisher = {APS}
}

@article{capua2006,
  title={Direct observation of spectroscopic inhomogeneities on La0. 7Sr0. 3MnO3 thin films by scanning tunnelling spectroscopy},
  author={Capua, R Di and Perroni, Carmine Antonio and Cataudella, Vittorio and Granozio, F Miletto and Perna, Paolo and Salluzzo, M and Uccio, U Scotti di and Vaglio, Ruggero},
  journal={Journal of Physics: Condensed Matter},
  volume={18},
  number={35},
  pages={8195--8204},
  year={2006}
}

@article{diez2022,
  title={Interfacial exchange phenomena driven by ferromagnetic domains},
  author={D{\'\i}ez, Jos{\'e} Manuel and Cu{\~n}ado, Jos{\'e} Luis F and Lapa, Pavel and Sol{\'\i}s, Ra{\'u}l and Arnay, Ic{\'\i}ar and Pedraz, Patricia and Perna, Paolo and Bollero, Alberto and Miranda, Rodolfo and Schuller, Ivan K and others},
  journal={Advanced Materials Interfaces},
  volume={9},
  number={21},
  pages={2200331},
  year={2022},
  publisher={Wiley Online Library}
}

@article{ali2007,
  title={Exchange bias using a spin glass},
  author={Ali, Mannan and Adie, Patrick and Marrows, Christopher H and Greig, Denis and Hickey, Bryan J and Stamps, Robert L},
  journal={Nature Materials},
  volume={6},
  number={1},
  pages={70--75},
  year={2007},
  publisher={Nature Publishing Group UK London}
}

@article{saghayezhian2019a,
  author    = {Saghayezhian, H. and Karki, S. and Marasini, J. and Zhang, J. and Ward, T. Z. and Plummer, E. W. and Zhang, J.},
  title     = {Spontaneous magnetic reversal in $\mathrm{La}_{2/3}\mathrm{Sr}_{1/3}\mathrm{MnO}_3$ heterostructures},
  journal   = {Physical Review Materials},
  volume    = {3},
  number    = {10},
  pages     = {104407},
  year      = {2019},
  publisher = {APS}
}

@article{mottaghi2018a,
  author    = {Mottaghi, N. and Trappen, R. and Kumari, S. and Huang, C.-Y. and Yousefi, S. and Seehra, M. S. and Holcomb, M. B.},
  title     = {Negative magnetic remanence and its mechanism in $\mathrm{La}_{0.7}\mathrm{Sr}_{0.3}\mathrm{MnO}_3$ thin films},
  journal   = {Journal of Physics: Condensed Matter},
  volume    = {30},
  number    = {45},
  pages     = {455802},
  year      = {2018},
  publisher = {IOP Publishing}
}

@incollection{wisniewski2017,
  title={Exchange-bias effect in bulk perovskite manganites},
  author={Wisniewski, A and Fita, I and Puzniak, R and Markovich, V},
  booktitle={Exchange Bias},
  pages={275--299},
  year={2017},
  publisher={CRC Press}
}

\end{document}